\documentclass[12pt]{iopart}
\usepackage{iopams}
\usepackage[bookmarks=true, pdfnewwindow=true]{hyperref}
\usepackage{graphicx}
\usepackage{graphics}
\usepackage[dvips]{color}
\usepackage{pdftexcmds}

\begin{document}

\title[Flexoelectric fluid membranes in electric field] {Flexoelectric fluid membranes in electric field. Shape equations and exact solutions}

\author{Galin S. Valchev and Vassil M. Vassilev }
\address{Institute of Mechanics, Bulgarian Academy of Sciences, \\ Acad. G. Bonchev St., Building 4, 1113 Sofia, Bulgaria}
\ead{gvalchev@imbm.bas.bg}
\vspace{10pt}

\begin{abstract}
The shape equation for an axisymmetric fluid membrane is derived, assuming action of an uniform external electric field. The flexoelectric contribution to the free energy of the membrane, stemming from the latter is accounted within the theory by Steigmann and Agrawal. Additionally, we have introduced, in the aforementioned functional, another term associated with a curvature induced membrane polarization, as the latter was first hypothesized by A. Petrov. Some exact Naito-type solutions of the studied equation are given, with the free parameters linked to the model ones. 
\end{abstract}

\vspace{2pc}

\section{Introduction} \label{Sec.1}

The plasma membrane is a complex organelle that encapsulates the contents of the cell. Phospholipids are among the most abundant molecules comprising this structure. They consist of hydrophobic fatty acid "tails" and a hydrophilic polar phosphate-containing "head" attached to either glycerol or sphingosine, thus resulting in fat- as well as water-soluble regions, respectively. Dissolved in water, the pure phospholipid phase undergoes a spontaneous process of self-assembly into bilayer structures, in order to isolate the tails from the aqueous medium, consequently reaching the system thermodynamic ground state where the free energy is minimal.

Since a pure phospholipid bilayer membrane is modeled as a liquid crystal, its state depends on the temperature. A variety of physical techniques have shown that such a structure experience a transition from a crystalline to a liquid crystalline form at a temperature dependent upon the presence and type of unsaturation in the fatty acid residues \cite{Chapman:1967}.

The theoretical study of the piezoelectric effect in liquid crystals by Meyer \cite{Meyer1969}, showed an analogous linear coupling between the electric
polarization and the curvature strain of such entities, designated as \textit{flexoelectric effect}. The piling experimental research since 1969, have shown that flexoelectricity is an omnipresent property exhibited by all dielectric materials, in comparison to piezoelectricity. In particular, it was observed by incident that a fluctuating lipid bilayer under oscillating pressure gradient, generates alternating current \cite{PassSol:1973,OchsBur:1974}. Shortly after these findings it was realised by A. Petrov and co-workers that indeed the periodically changing bilayer curvature is directly linked to the oscillating flexoelectric polarization \cite{Petrov1975,PetDer:1976,DerPetPav:1981}.

In the present theoretical letter, we derive the shape equation of an axisymmetric fluid membrane immersed in uniform electric field, and discuss some exact solutions of that equation. The contributions of the flexoelectric effect to the free energy of the membrane are described both by the recent theory by Steigmann and Agrawal \cite{Steigmann2016} and the pioneering phenomenological expression introduced by Petrov (see p. 295 in \cite{Petrov1999} as well as pp. 66--68 in \cite{Ou-Yang:1999}).
\section{The Model}\label{model}
The foundation of the current theoretical understanding of the shapes and mechanical response of fluid membranes to different types of excitations can be traced more than thirty years back (see, e.g., \cite{ Ou-Yang:1999, Lipowsky:1995, Seifert:1997}) to the works by Canham \cite{Canham:1970} and Helfrich \cite{Helfrich:1973} in which the first one of the so-called curvature models was introduced and developed.
In all models of this kind, the vesicle's membrane is regarded as a two-dimensional surface $\mathcal{S}$ embedded in the three-dimensional Euclidean space and assumed to exhibit purely elastic behaviour (elastic bending without stretching) described in terms of its mean $H$ and Gaussian $K$ curvatures and two material constants associated with the bending rigidity of the membrane.

Within the Helfrich spontaneous curvature model \cite{Helfrich:1973}, a two-dimensional surface $\mathcal{S}$ representing a fluid membrane in equilibrium is assumed to provide a local extremum of the elastic bending energy functional
\begin{equation}\label{Eb}
\mathcal{F}_{b}=\frac{k_{c}}{2}\int_{\mathcal{S}}\left( 2H - c_0\right)^{2}{
	\mathrm{d}}A+k_G\int_{\mathcal{S}}K{\mathrm{d}}A,
\end{equation}
under the constraints of fixed total membrane area $A$ and enclosed volume $V$ (if the membrane is subjected to a uniform hydrostatic pressure $P$). This functional is also called curvature or shape energy. Here $c_0$, $k_{c}$ and $k_G$ are three real constants representing the spontaneous curvature (a constant introduced by Helfrich to reflect a possible asymmetry of the membrane or its environment, chemical structures of the molecules, salinity and temperature in the vesicle), bending and Gaussian rigidity of the membrane, respectively.

Using two Lagrange multipliers $\Lambda $ and $P$ to take into account the aforementioned constraints, this yields the functional
\begin{equation}\label{Ebc}
\mathcal{F}_{bc}=\frac{k_{c}}{2}\int_{\mathcal{S}}\left( 2H-c_0\right) ^{2}{
	\mathrm{d}}A+k_G\int_{\mathcal{S}}K{\mathrm{d}}A+\Lambda \int_{\mathcal{S}}{\mathrm{d}}A+P\int {\mathrm{d}}V.
\end{equation}
The Lagrange multiplier $\Lambda$ corresponds to the constraint of fixed total area and can be interpreted as the tensile stress or chemical potential associated with the number of the lipid molecules located at the membrane surface, while the pressure $P$ appearing as another Lagrange multiplier corresponds to the constraint of fixed enclosed volume $V$.

A phenomenological expression for a curvature induced membrane  polarization was first given in \cite{Petrov1975}. Accordingly, the flexoelectric effect is accounted for by including the following term in the energy functional (\ref{Ebc})
\begin{equation}\label{Efl}
\mathcal{F}_{fl}=C_{fl} \int_{\mathcal{S}} H \mathrm{d} A,
\end{equation}
(see, e.g., \cite{Petrov1999} and pp. 66--68 in \cite{Ou-Yang:1999}) where $C_{fl}$ is termed flexoelectric coefficient.

In the presence of an external electric field $\mathbf{ E}_{ext}$ another flexoelectric contribution \cite{Abtahi2020} (see also \cite{Steigmann2016}) adds up to the membrane energy
\begin{equation}\label{Eef}
\mathcal{F}_{ef}=\frac{1}{2D}\int_{\mathcal{S}} \left[ \left( \mathbf{E}_{ext} \cdot \mathbf{n} \right) ^2+\mathbf{ E}_{ext} \cdot \mathbf{ E}_{ext} \right] \mathrm{d} A,
\end{equation}
where $\mathbf{n}$ is the (outward) normal vector of the membrane surface, $D$ is a  material constant characterizing the strength of the flexoelectric effect and the dot $ (\cdot)$ denotes the usual scalar product between vectors. In deriving Eq. (\ref{Eef}) Steigmann and Agrawal assumed that the polarization vector is essentially tangential to the membrane surface. To some extent this simplification is supported by molecular dynamics simulations \cite{WMC2011,TA2021}. Explicitly, the material constant is given by
\begin{equation}\label{Dconst}
 D=\chi_{\bot}-\frac{c_{2}^{2}}{k_{3}},
\end{equation}
where $\chi_{\bot}$ is the electric polarisability, exhibited by the membrane when $\mathbf{E}_{ext}$ is applied, $c_{2}$ is a constant proportional to the strength of flexoelectric coupling between the polarisation and tangential, to the membrane surface vectors, $k_{3}$ is the bend modulus of the Frank energy of nematic liquid crystals. From Eq. (\ref{Dconst}) it is seen that the sign of $D$ depends on the "strength" of the flexoelectric effect, encoded by $c_{2}$. For weak polarization-tangential coupling one has $c_{2}^{2}< k_{3}\chi_{\bot}$, i.e. $D>0$ and vice versa when the coupling is strong.

Summarizing the aforementioned considerations, the energy $\mathcal{W}$ of a flexo\-electric fluid membrane in electric field may be expressed in the form
\begin{equation} \label{FLEFM}
	\mathcal{W}=\mathcal{F}_{bc}+\mathcal{F}_{fl}+\mathcal{F}_{ef}.
\end{equation}
The stationarity of this energy leads to the corresponding Euler-Lagrange equations determining the equilibrium shapes of a flexo\-electric fluid membrane in electric field, that is the shape equations. The variational procedure used for that purpose and the derived shape equations are given in \cite[Sections 4 and 5]{Steigmann2016}. The special case of axisymmetric states is considered in \cite[Section 6]{Steigmann2016} and in \cite{Abtahi2020}.

An alternative derivation of the shape equations for axisymmetric flexoelectric fluid membranes will be given in the next section.

\section{Shape equations for axisymmetric membranes} \label{Sec.3}

Let $ X,Y,Z $ denote the axes of a right-handed rectangular Cartesian coordinate system $ \{x,y,z \}$ in the three-dimensional Euclidean space $\mathbb{R}^{3}$.
Without loss of generality, the components $ x,y,z $  of the position vector $ \mathbf{x} $ of an axially symmetric surface $\mathcal{S}$ immersed in $\mathbb{R}^{3}$ can be given in the form
\begin{equation} \label{PV}
	\mathbf{x}(u, v )=\left[
	\begin{array}{c}
		x(u, v ) \\
		y(u, v ) \\
		z(u, v )
	\end{array}
	\right] =\left[
	\begin{array}{c}
		r(u)\cos v  \\
		r(u)\sin v  \\
		h(u)
	\end{array}
	\right], \quad
	u \in \Omega \subseteq {\mathbb R}, \quad
	v \in  [0,2 \pi],
\end{equation}
where the functions $ r(u) $ and $ h(u) $ are supposed to have as many derivatives as may be required on the domain $ \Omega $.
Such a surface can be thought of as obtained by revolving around the $OZ$-axis a plane curve $\Gamma: u \mapsto [r(u), h(u)]$ laying in the $XOZ$-plane.

In this notations, the first and second fundamental tensors of the
surface $\mathcal{S}$ read
\begin{equation} \label{gb}
	g_{\alpha \beta }=
	\left(\begin{array}{cc} r_{u}^{2}+h_{u}^{2} & 0 \\
		0 & r^{2}\\
	\end{array}
	\right), \quad
	b_{\alpha \beta }=
	\frac{1}{\sqrt{r_{u}^{2}+h_{u}^{2}}}\left(\begin{array}{cc} r_{u}h_{uu}-h_{u}r_{uu} & 0 \\
		0 & r h_{u}\\
	\end{array}
	\right)
\end{equation}
respectively, and
\begin{equation} \label{g}
	g=\det (g_{\alpha \beta })=r^{2} \left(r_{u}^{2}+h_{u}^{2}\right), \quad
	d A=\sqrt{g}\, d u d v=r \sqrt{r_{u}^{2}+h_{u}^{2}}\, d u d v.
\end{equation}
Consequently, the mean and Gaussian curvatures of the regarded surface take the following forms
\begin{equation} \label{H-K}
	H=\frac{1}{2}\frac{r\left( r_{u}h_{uu}-h_{u}r_{uu}\right) +h_{u}\left(
		r_{u}^{2}+h_{u}^{2}\right) }{r\left( r_{u}^{2}+h_{u}^{2}\right) ^{3/2}}, \quad
	K=\frac{h_{u}\left( r_{u}h_{uu}-h_{u}r_{uu}\right) }{r\left(
		r_{u}^{2}+h_{u}^{2}\right) ^{2}} \cdot
\end{equation}
Here and in what follows, subscripts $u$ and $v$ denote derivatives with respect to the corresponding variables.

For a constant electric field of magnitude $\epsilon $ pointing in the vertical $Z$-direction,
on account of Eqs. (\ref{Ebc}), (\ref{Efl}), (\ref{Eef}) (\ref{g}) and (\ref{H-K}), the energy $\mathcal{W}$ of an axisymmetric flexoelectric fluid membrane represented by the surface $\mathcal{S}$ reads
\begin{equation} \label{WEn}
	\mathcal{W}=2\pi \int_\Omega \mathcal{L}d u, \quad \mathcal{L}=\mathcal{L}_{bc}+\mathcal{L}_{fl}+\mathcal{L}_{ef},
\end{equation}
where
\begin{eqnarray}
	\mathcal{L}_{bc} &=&\frac{k_{c}}{2}\left[ \frac{r\left(
		r_{u}h_{uu}-h_{u}r_{uu}\right) +h_{u}\left( r_{u}^{2}+h_{uu}^{2}\right) }{%
		r\left( r_{u}^{2}+h_{u}^{2}\right) ^{3/2}}-c_{0}\right] ^{2}r\sqrt{%
		r_{u}^{2}+h_{u}^{2}}  \nonumber \\
	&+&k_{G}\frac{h_{u}\left( r_{u}h_{uu}-h_{u}r_{uu}\right) }{\left(
		r_{u}^{2}+h_{u}^{2}\right) ^{3/2}}
	+\Lambda r\sqrt{r_{u}^{2}+h_{u}^{2}}-\frac{1}{3}Pr\left( rh_{u}-hr_{u}\right),
	\label{Lbc}
\end{eqnarray}
\begin{equation}\label{Lfl}
	\mathcal{L}_{fl}=C_{fl}  \frac{1}{2}\frac{r\left( r_{u}h_{uu}-h_{u}r_{uu}\right) +h_{u}\left(r_{u}^{2}+h_{u}^{2}\right) }{r_{u}^{2}+h_{u}^{2}},
\end{equation}
\begin{equation}\label{Lef}
	\mathcal{L}_{fl}= \frac{\epsilon^2}{2 D}\frac{h_u^2}{r_{u}^{2}+h_{u}^{2}}.
\end{equation}


The application of the Euler operators
\[
E_r=\frac{\partial }{\partial r}-D_{u }\frac{\partial }{\partial
	r_{u }}+D_{t }D_{u }\frac{\partial }{\partial r_{uu }}-\cdots
\]
\[
E_h=\frac{\partial }{\partial h}-D_{u }\frac{\partial }{\partial
	h_{u }}+D_{u }D_{u }\frac{\partial }{\partial h_{uu }}-\cdots
\]
where
\[
D_{u }=\frac{\partial }{\partial u}
+r_{u}\frac{\partial }{\partial r}
+h_{u}\frac{\partial }{\partial h}
+r_{uu }\frac{\partial }{\partial r_{u }}
+h_{uu }\frac{\partial }{\partial h_{u }}
+\cdots
\]
is the total differentiation operator, on the Lagrangian density $\mathcal{L}$ of the functional (\ref{WEn}) leads to the Euler-Lagrange equations $E_r \mathcal{L}=0$ and $E_h \mathcal{L}=0$. However, it turned out that $ r_u E_r \mathcal{L}\equiv-h_u E_h \mathcal{L} $ and hence we face an abnormal variational problem (see Sec. 5.3 in Ref. \cite{Olver2003}) having a single equation, say $E_r \mathcal{L}=0$,
instead of a system of two Euler-Lagrange equations for two dependent variables $ r(u) $ and $ h(u) $ as expected normally. Thus, we have got an under-determined system and to complete it we may add another equation of our own choice in addition to equation $E_r \mathcal{L}=0$.

In these notes, we will consider a complementary equation of the form
\begin{equation}
	r_{u}^{2}+h_{u}^{2}=1.
\end{equation}
This means that $ u $ is the arc length $ s $ along the profile curve $\Gamma$.
Then, in terms of the tangent angle $ \psi $ we have
\begin{equation} \label{cs}
\frac{\mathrm{d}r }{\mathrm{d}s}=\cos \psi, \qquad
\frac{\mathrm{d}h }{\mathrm{d}s}=\sin  \psi
\end{equation}
and can rewrite equation $E_r \mathcal{L}=0$ in the form
\begin{eqnarray}
	\frac{\mathrm{d}^{3}\psi }{\mathrm{d}s^{3}} &+&\frac{2\cos \psi }{r} \frac{\mathrm{d}^{2}\psi }{\mathrm{d}s^{2}}+\frac{1}{2} \left(\frac{\mathrm{d}\psi}{\mathrm{d}s}\right) ^{3}- \frac{3\sin\psi}{2r}\left( \frac{\mathrm{d}\psi}{\mathrm{d}s}\right) ^{2} \nonumber \\
	&+&\left(\frac{1-3\cos ^{2}\psi}{2r^{2}}+ \frac{2c_{0}-c_{fl}}{r}\sin \psi-\frac{1}{2}c_{0}^{2}-\lambda \right) \frac{\mathrm{d}\psi }{\mathrm{d}s} \nonumber \\
	&+&\varepsilon \left( \frac{3+\cos \left( 2\psi \right) }{r}\sin \psi
	+\left[ 1+3\cos \left( 2\psi \right) \right] \frac{\mathrm{d}\psi }{\mathrm{d%
		}s}\right)  \nonumber\\
	&+& \left( \frac{1+\cos ^{2}\psi }{2r^{2}}-\frac{1}{2}c_{0}^{2}-\lambda
	\right) \frac{1}{r}\sin \psi = p.
	\label{ASSEq}
\end{eqnarray}
where $p=P/k_c$, $\lambda =\Lambda /k_c$, $c_{fl}=C_{fl}/k_c$ and $\varepsilon=\epsilon^2 /(4Dk_c)$.
We obtain also the following conservation law
\begin{eqnarray}
	\frac{\mathrm{d}^{2}\psi }{\mathrm{d}s^{2}} &+&\frac{1}{2}\tan \psi \left(\frac{\mathrm{d}\psi }{\mathrm{d}s}\right) ^{2}+\frac{\cos \psi }{r}\frac{\mathrm{d}\psi }{\mathrm{d}s}
	-\frac{\left( 1+\cos^2 \psi  \right) \tan \psi }{2 r^{2} }
	\nonumber \\
	&+&\left( 2c_{0}-c_{fl}\right) \frac{\sin ^{2}\psi }{2r\cos \psi }+\varepsilon \left[ \sin \left( 2\psi \right) +2\tan \psi \right]  \nonumber \\
	&-&	\frac{1}{2}\left( c_{0}^{2}-2\lambda \right) \tan \psi
	+ \frac{p x }{6 \cos \psi}   \nonumber \\
	&=&\frac{c}{r\cos \psi },
	\label{CLASs}
\end{eqnarray}
which holds on the solutions of Eq. (\ref{ASSEq}). Here, $c$ is the constant of integration.
Let us remark that if $c_{fl}=0$ and $\varepsilon=0$, then we have the well-known shape equation and its first integral.

\section{Some exact solutions}

In  \cite{Naito1995}, Naito and co-workers discovered that the ''standard'' shape equation for axisymmetric fluid membranes (i.e., Eq. (\ref{ASSEq}) with $c_{fl}=0$ and $\varepsilon=0$) admits exact solutions of the form
\begin{equation}
\psi =\arcsin \left(ar+b+dr^{-1}\right)
\label{sol}
\end{equation}
provided that $a$, $b$ and $d$ are real constants which meet certain conditions. Substituting expression (\ref{sol}) into Eq. (\ref{ASSEq}) and using Eqs. (\ref{cs}) we obtain the following three types of exact solutions of Eq. (\ref{ASSEq}) in the cases when at least one of the constants  $c_{fl}=0$ or $\varepsilon=0$ is nonzero.

\textbf{Case 1.} Let $c_{fl} \neq 0$, $\varepsilon \neq 0$, $p=0$, then
\begin{equation}\label{ExS1}
\psi =\arcsin \left( \frac{2 c_0-c_{fl}}{\varepsilon r} \right)
\end{equation}
is an exact solution of Eq. (\ref{ASSEq}).

\textbf{Case 2.} Let $c_{fl} \neq 0$, $\varepsilon= 0$, $\lambda=-c_0^2/2$, $p=-\left(2 c_0-c_{fl} \right)^3/16$, then
\begin{equation}\label{ExS2}
	\psi =\arcsin \left[ \frac{d}{r} +\frac{1}{4} \left(2 c_0-c_{fl} \right) r\right]
\end{equation}
is an exact solution of Eq. (\ref{ASSEq}).

\textbf{Case 3.} Let $c_{fl}= 0$, $\varepsilon \neq 0$, $p=0$, then
\begin{equation}\label{ExS3}
	\psi =\arcsin \left( \frac{2 c_0}{\varepsilon r} \right)
\end{equation}
is an exact solution of Eq. (\ref{ASSEq}).

In all the cases listed above, in order to find the parametric representation of the corresponding profile curves one should find the solutions of the Eqs. (\ref{cs}) in which the tangent angle $\psi$ is given by the expressions (\ref{ExS1}), (\ref{ExS2}) or (\ref{ExS3}), respectively.


\section*{Acknowledgements}

The authors gratefully acknowledge the financial support via contract K$\Pi$-06-H22/2 with the Bulgarian National Science Fund.

\section*{References}

\providecommand{\newblock}{}

\end{document}